**Quel jour naît-on le plus en France ?**

**Sur les dix dernières années, de 2015 à 2024, le jour ayant en moyenne le maximum de naissances est le 20 juillet, tandis que Noël est le jour en ayant le minimum. En dehors de fin juillet-début août, les jours avec le plus de naissances se situent fin septembre, ce qui correspond à une conception au moment des fêtes de fin d'année. Les naissances sont moins fréquentes les jours fériés et les week-ends.**

**Depuis les années 1970, la saison des bébés s'est décalée du printemps à l'été. De 1975 à 2010, les vagues de chaleur ont le plus souvent été suivies d'un déficit de naissances 9 mois plus tard en avril ou mai ; ce déficit est moins visible depuis, dans un contexte où les vagues de chaleur se multiplient. Par ailleurs, le pic des naissances en avril des mères professeures des écoles a fortement diminué depuis les années 1970.**

Nathalie Blanpain (Insee), Insee focus n°360, septembre 2025

### +9 % le 20 juillet et -22 % de naissances à Noël

De 2015 à 2024, le jour ayant le maximum de naissances est le 20 juillet, avec un nombre moyen de naissances de 2 210, contre 2 030 sur l'ensemble de la période, soit une surnatalité de +9 % **(figure 1a, méthode)**. Le 20 juillet est un jour d'été, soit la saison avec le plus de naissances. Étant donné que 30 % des femmes ont accouché avant 39 semaines d'aménorrhées (SA) et 57 % avant 40 SA [Cinelli et al., 2022], la durée médiane d'une grossesse peut être estimée à 264 jours. Une naissance le 20 juillet correspond donc à une conception vers le 29 octobre, soit pendant les vacances de la Toussaint. Or, les vacances scolaires semblent davantage propices aux conceptions. En effet, un surplus de naissances par rapport aux jours adjacents a lieu mi-mai, les deuxièmes quinzaines de juillet et de septembre, ainsi que mi-novembre, ce qui correspond à des conceptions mi-août, lors des vacances de la Toussaint, de Noël et d'hiver.

Parmi les dix jours concentrant le plus de naissances, sept se situent fin juillet ou début août. Les autres jours les plus fréquents se situent fin septembre, ce qui correspond à une conception au moment des fêtes de fin d'année. C'est une période de vacances, qui peut être, de plus, associée à une moindre vigilance contraceptive en raison des festivités (oublis de pilule, retard dans sa prise [Rohrbasser, Régnier-Loilier, 2011]).

À l'opposé, le jour ayant le minimum de naissances est le 25 décembre, avec 1 600 naissances en moyenne ce jour-là, soit une sous-natalité de -22 %. Il s'agit d'un jour d'hiver, qui est de plus férié. Or, les naissances lors des jours fériés sont moins fréquentes : par exemple, -17 % le 1er janvier, -11 % le 1er mai, -9 % le 1er novembre et le 8 mai, -7 % le 11 novembre. Ceci pourrait s'expliquer par un moindre nombre d'accouchements programmés lors de ces jours de repos habituels : 7 % des naissances ont lieu par césarienne programmée et 26 % suite à un accouchement déclenché, qui peut être programmé ou non [Cinelli et al., 2022].

La sous-natalité est aussi accentuée lors de certains jours non fériés par rapport à la période adjacente : le 29 février (-10 %, **méthode**), ce qui pourrait s'expliquer par un souhait des parents d'éviter ce jour, sans date d'anniversaire trois années sur quatre ; les jours proches de Noël et du nouvel an comme le 31 décembre (-7 %) ; et enfin les jours du 1 au 5 septembre, qui correspondent à la fin des vacances scolaire et à la rentrée, ce qui peut engendrer une baisse des accouchements programmés, pour des difficultés de garde d'enfants ou de présence du personnel médical. D'ailleurs, le creux observé est un peu plus accentué pour les mères âgées de 35 ans ou plus qui ont plus souvent déjà un enfant et sont davantage concernées par les césariennes programmées [Baubeau, Buisson, 2003], que pour celles âgées de 24 ans ou moins.

Figure 1a - Écart au nombre moyen de naissances par jour de 2015 à 2024

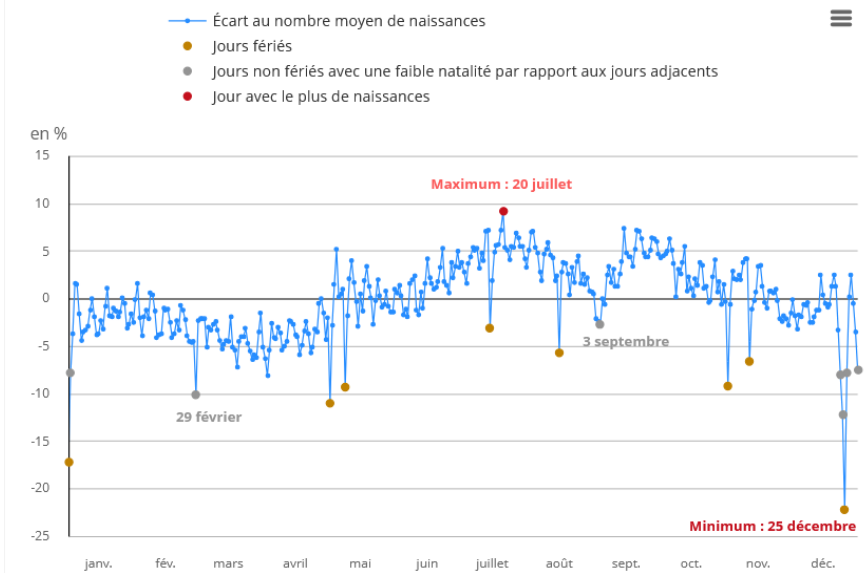

Notes : Seuls les jours fériés à date fixe peuvent être représentés. Le calcul pour le 29 février tient compte du fait que ce jour n'est présent que trois fois sur la période (**méthodes**).
Lecture : De 2015 à 2024, le nombre moyen de naissances le 25 décembre était inférieur de 22,2 % au nombre moyen de naissances par jour.
Champ : Naissances ayant eu lieu en France.
Source : Insee, statistiques d'état civil.

## -13 % de naissances le dimanche et -9 % le samedi

Les jours de la semaine ayant le maximum de naissances sont le mardi et le vendredi, avec 2 150 naissances en moyenne, soit une surnatalité de +6 % ; le jour avec le minimum de naissances est le dimanche, avec 1 760 naissances (-13 %), suivi par le samedi avec 1 840 naissances (-9 %, **figure 2**). Comme pour les jours fériés, la sous-natalité le week-end pourrait s'expliquer par un nombre moins élevé d'accouchements programmés ces jours-là. D'ailleurs, la sous-natalité le week-end s'accentue avec l'âge de la mère ; par exemple, chez les mères âgées de 40 ans ou plus, qui sont davantage concernées par les césariennes programmées, les naissances se produisent moins fréquemment le dimanche (-23 %) que chez les mères âgées de 24 ans ou moins (-8 %).

Figure 2 - Écart au nombre moyen de naissances selon le jour de la semaine et l'âge de la mère de 2015 à 2024

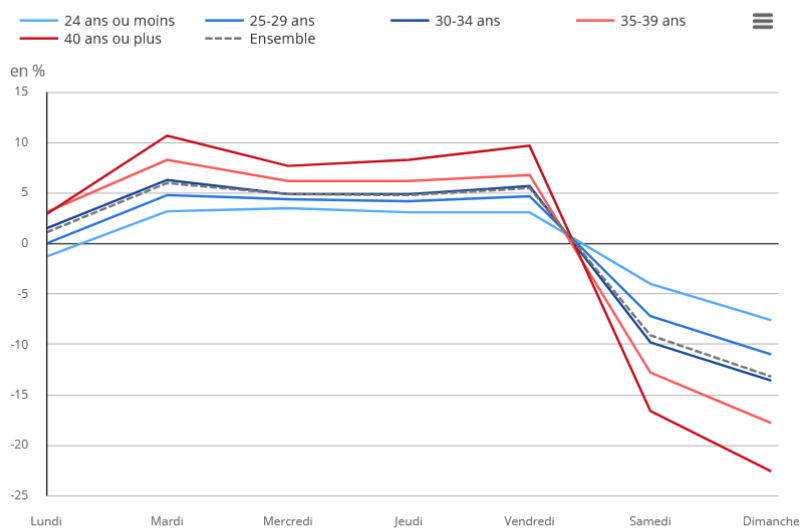

Lecture : De 2015 à 2024, pour les mères âgés de 40 ans ou plus, le nombre moyen de naissances le dimanche était inférieur de 22,6 % au nombre moyen de naissances par jour.
Champ : Naissances ayant eu lieu en France.
Source : Insee, statistiques d'état civil.

Ce déficit de naissances le week-end n'est pas un phénomène nouveau. De 1975 à 1995, il s'est très nettement accentué, de -4 % à -19 % (figure 3), probablement en lien avec l'augmentation des accouchements programmés. Ce déficit est resté quasi stable de 1995 à 2002. Il se réduit depuis 2002 pour s'établir à -11 % en 2024.

Figure 3 - Ecart au nombre moyen de naissances le week-end de 1975 à 2024

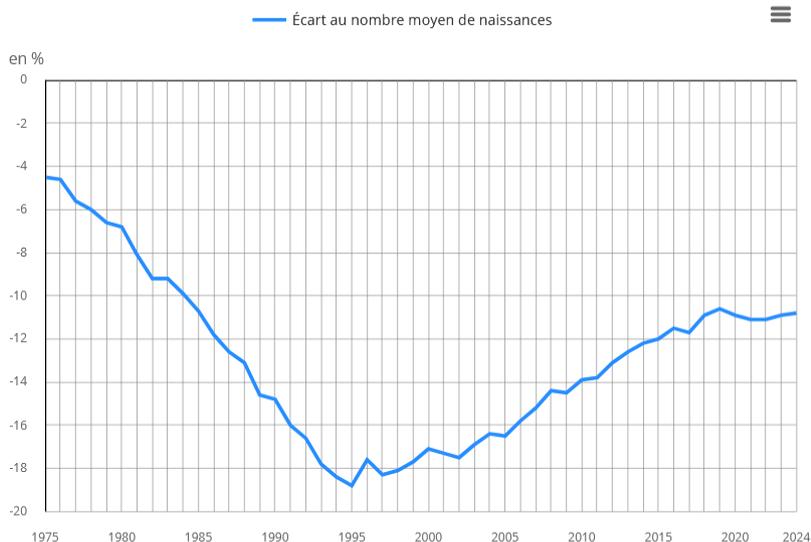

Lecture : En 2024, le nombre moyen de naissances le week-end était inférieur de 10,8 % au nombre moyen de naissances par jour.
Champ : Naissances ayant eu lieu en France métropolitaine jusqu'en 1997, en France hors Mayotte de 1998 à 2013, en France depuis 2014.
Source : Insee, statistiques d'état civil.

### Depuis les années 1970, la saison des bébés s'est décalée du printemps à l'été

De 2015 à 2024, 2 120 personnes sont nées en moyenne par jour en juillet, contre 2 030 sur l'ensemble de la période, soit +5 % (figure 4). Les mois d'été et du début de l'automne (de juillet à octobre) sont les mois avec le plus de naissances (+2 % à +5 %). À l'inverse, les mois de mars et d'avril sont les mois avec le moins de naissances (-4 %).

Cette saisonnalité des naissances varie peu selon l'âge de la mère. Seules les mères âgées de 40 ans ou plus se distinguent avec un déficit des naissances en mai (-4 %, contre 0 % en moyenne), ce qui pourrait s'expliquer par une activité réduite des centres de procréation médicalement assistée en août.

Depuis les années 1970, la période où les naissances sont les plus nombreuses s'est progressivement décalée du printemps et début de l'été, vers l'été et le début de l'automne. Dans les années 1970, le pic des naissances au printemps pouvait s'expliquer notamment par les congés estivaux et la part élevée de mariages célébrés en été, ainsi que par la préférence des parents d'avoir un enfant à cette période [Papon, 2020]. Sur la période récente, la part des naissances hors mariage est élevée (58 % en 2024), aussi la saisonnalité des mariages explique moins qu'auparavant celle des naissances. On ne dispose pas de données récentes sur le mois préféré des parents pour la naissance de leur enfant ; il est possible que cette préférence soit désormais en été, une partie des parents pouvant choisir le moment de la naissance de leur enfant : la probabilité de concevoir un enfant au cours d'un cycle est en moyenne de 20 % à 25 % vers 20-30 ans [Leridon, 2010].

L'amplitude de la saisonnalité mensuelle des naissances, c'est-à-dire l'écart de natalité entre les mois où elle est la plus faible et la plus forte, a baissé de 1975-1984 (15 points) à 1985-1994 (9 points) ; depuis cette date, elle reste constante.

Figure 4 - Écart au nombre moyen de naissances par jour selon le mois et la période

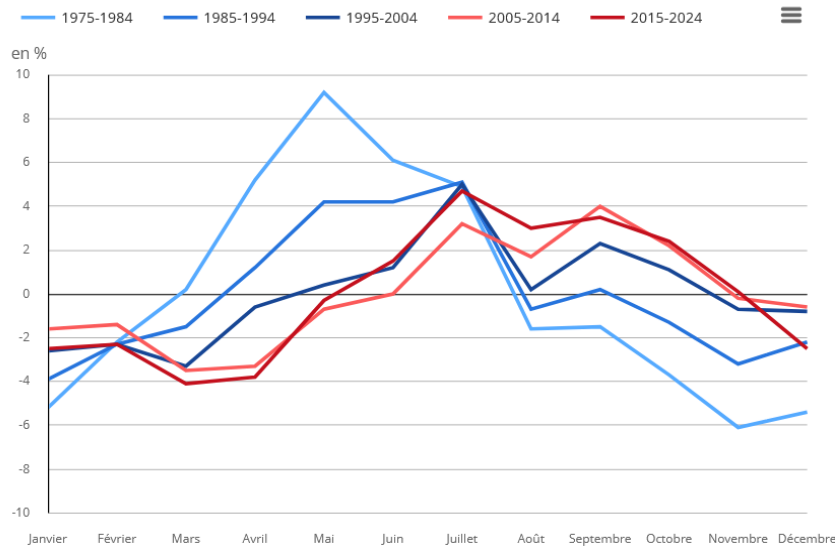

Lecture : De 1975 à 1984, le nombre moyen de naissances par jour en mai était supérieur de 9,2 % au nombre moyen de naissances par jour.
Champ : Naissances ayant eu lieu en France métropolitaine jusqu'en 1997, en France hors Mayotte de 1998 à 2013, en France depuis 2014.
Source : Insee, statistiques d'état civil.

**Un déficit de naissances 9 mois après les vagues de chaleur plus faible et moins systématique depuis 2010**

De 1975 à 2010, les **vagues de chaleur** ont généralement été suivies d'un déficit marqué de naissances 9 mois plus tard, en avril ou mai **(figure 5)**. Par exemple, 9 mois après la **canicule** d'août 2003, le nombre moyen de naissances par jour en mai 2004 est inférieur de -6 % à la moyenne de l'année, contre +2 % en mai 2003 et -1 % en mai 2005. Jusqu'en 2010, les périodes de températures très élevées semblaient moins propices aux conceptions. En revanche, depuis 2010, le déficit de naissances 9 mois après les vagues de chaleur est plus faible et moins systématique, par rapport aux années sans vague. Depuis 2010, l'écart de natalité entre les mois avec et sans vague de chaleur 9 mois plus tôt est faible (-1,3 point en moyenne). La natalité lors des mois avec vague est donc peu éloignée de celle lors des mois sans vague. A titre de comparaison, avant 2010, l'écart est en moyenne de -3,2 points. Il est possible que la multiplication des vagues de chaleur depuis 2010 ait entraîné une adaptation des comportements ou des habitats (moindre baisse de l'activité sexuelle).

Figure 5a - Ecart au nombre moyen de naissances par jour en avril selon l'année

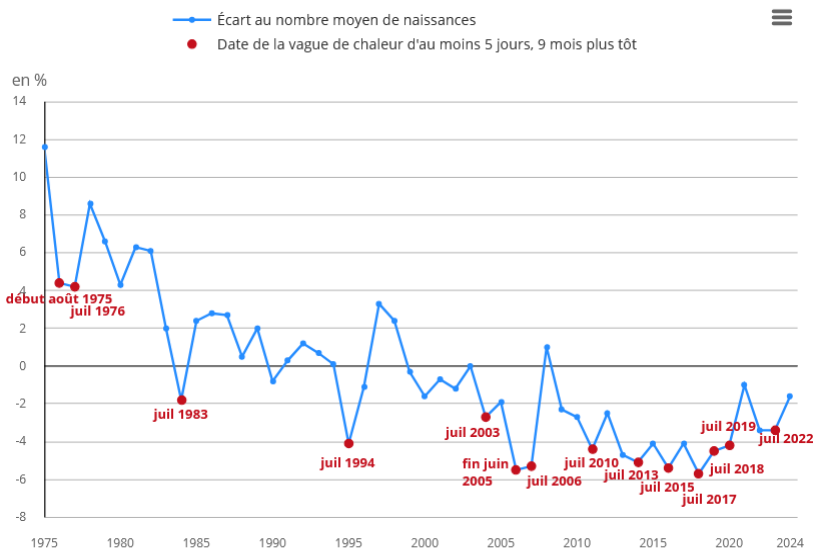

Lecture : En avril 1984, neuf mois après la vague de chaleur de juillet 1983, le nombre moyen de naissances par jour était inférieur de 1,8 % au nombre moyen de naissances par jour en 1984.
Champ : Naissances ayant eu lieu en France métropolitaine jusqu'en 1997, en France hors Mayotte de 1998 à 2013, en France depuis 2014.
Source : Insee, statistiques d'état civil.

Figure 5b - Écart au nombre moyen de naissances par jour en mai selon l'année

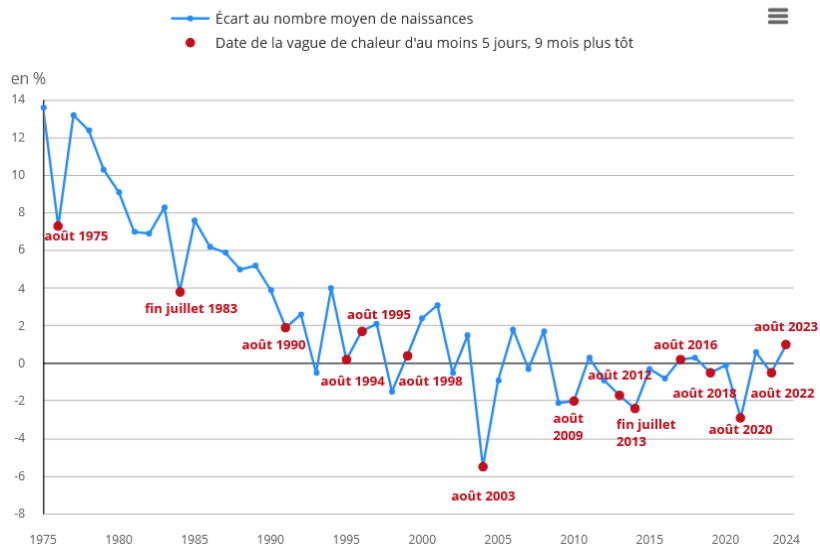

Lecture : En mai 2004, neuf mois après la vague de chaleur d'août 2003, le nombre moyen de naissances par jour était inférieur de 5,5 % au nombre moyen de naissances par jour en 2004.
Champ : Naissances ayant eu lieu en France métropolitaine jusqu'en 1997, en France hors Mayotte de 1998 à 2013, en France depuis 2014.
Source : Insee, statistiques d'état civil.

### Une saisonnalité des naissances différente dans les DOM

L'ampleur de la saisonnalité mensuelle des naissances et le moment du pic (en juillet) sont les mêmes quelles que soient les régions de France métropolitaine. En revanche, l'ampleur de la saisonnalité est plus forte lorsque la mère réside dans les DOM, avec un pic de +26 % à Mayotte, +14 % en Guadeloupe, +12 % en Martinique, +9 % en Guyane et à la Réunion, contre +5 % en France métropolitaine **(figure 6)**.

De plus, le pic de naissances se produit à des périodes différentes. Il a lieu en avril pour les mères résidant à Mayotte et à la Réunion, soit une conception en juillet, pendant les vacances scolaires. Pour Mayotte, cela pourrait s'expliquer également par un taux plus élevé de mariages religieux en juillet-août [Touzet, 2019]. Pour les mères résidant en Guadeloupe, en Guyane et en Martinique, la surnatalité se produit de septembre à décembre, soit une conception pendant la période de Noël ou celle du Carnaval (de janvier à mars aux Antilles et en Guyane). La période de Carnaval est davantage propice aux conceptions : de 2000 à 2011, un pic de naissances s'est produit 9 mois plus tard en Guadeloupe, sauf en 2009, une grève ayant empêché la tenue du Carnaval [Kadhel et al., 2017].

Figure 6 - Ecart au nombre moyen de naissances par jour selon le mois et le lieu de domicile de la mère de 2015 à 2024

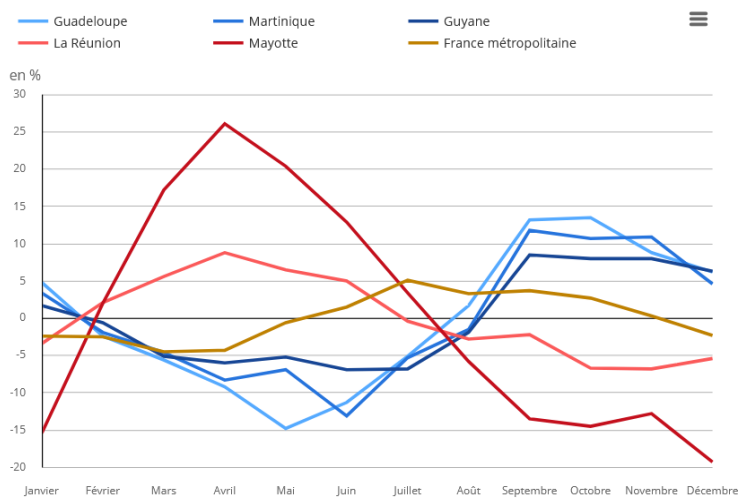

Lecture : De 2015 à 2024, pour les mères résidant à Mayotte, le nombre moyen de naissances par jour en avril était supérieur de 26,1 % au nombre moyen de naissances par jour.
Champ : Naissances ayant eu lieu en France et mères résidant en France.
Source : Insee, statistiques d'état civil.

**Le pic des naissances en avril des mères professeures des écoles a très nettement diminué depuis les années 1970**

De 2015 à 2024, le pic des naissances est en juillet quelle que soit la catégorie sociale (en sept modalités) des mères ; il est cependant moins prononcé pour les mères inactives ou de catégorie sociale inconnue (+3 %), alors qu'il est plus élevé pour les ouvrières et les cadres (+7 %). Les agricultrices accouchent moins souvent que les autres en avril (-8 %) et en mai (-5 %), ce qui correspond à des conceptions en juillet-août, période souvent intense de travail.

Les mères **professeures des écoles (ou assimilées)** se distinguent : leur pic d'accouchements a lieu en avril, ce qui correspond à une conception en juillet, pendant les vacances scolaires, moment de plus grande disponibilité des professeures. De plus, un accouchement en avril conduit à faire coïncider la fin du congé maternité avec le début des vacances scolaires d'été [Régnier-Loilier, Wiles-Portier, 2010].

L'ampleur de la saisonnalité des naissances des mères professeures des écoles a fortement diminué depuis les années 1970 **(figure 7)**. Le pic de surnatalité était de +39 % en 1975-1984, +14 % en 2005-2014, il n'est plus que de +6 % en 2015-2024. Cela pourrait refléter une évolution des préférences ou bien une plus grande difficulté des couples à concevoir un enfant au moment souhaité, du fait du recul de l'âge moyen à l'accouchement (de 27 ans en 1975 à 31 ans en 2024 pour l'ensemble des mères).

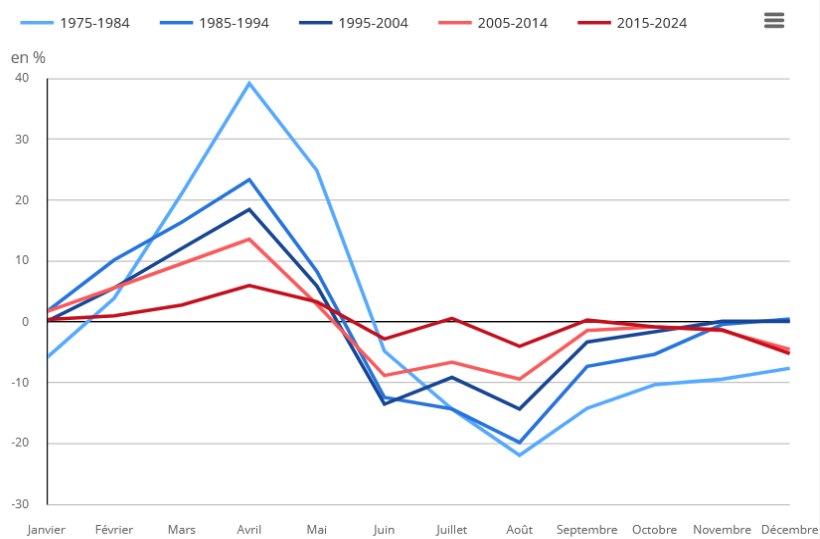

Figure 7 - Écart au nombre moyen de naissances par jour selon le mois et la période pour les mères professeures des écoles

Lecture : De 1975 à 1984, pour les mères professeures des écoles ou assimilées, le nombre moyen de naissances par jour en avril était supérieur de 39,2 % au nombre moyen de naissances par jour.
Champ : Naissances ayant eu lieu en France métropolitaine jusqu'en 1997, en France hors Mayotte de 1998 à 2013, en France depuis 2014.
Source : Insee, statistiques d'état civil.

**Sources**

Les **statistiques d'état civil** (dont les naissances) sont issues des bulletins transmis par les mairies à l'Insee. Le code civil oblige en effet à déclarer tout événement d'état civil à un officier d'état civil. L'Insee s'assure de l'exhaustivité et de la qualité des données avant de produire les fichiers statistiques d'état civil.

**Méthode**

La surnatalité ou la sous-natalité lors d'un mois $m$ est le rapport entre le nombre moyen des naissances par jour d'une population lors de ce mois $m$ sur une période donnée et le nombre moyen des naissances par jour de cette population sur l'ensemble de cette période.

La surnatalité ou la sous-natalité lors d'un jour *j* est le rapport entre le nombre moyen des naissances d'une population lors de ce jour *j* sur une période donnée et le nombre moyen des naissances par jour de cette population sur l'ensemble de cette période. On ne tient pas compte du fait que chaque jour comporte un nombre différent de jours de la semaine sur la période (par exemple, aucun dimanche le 20 juillet de 2015 à 2024). Une méthode a été testée pour corriger les effets des jours de la semaine **(figure 1b)**. Les résultats sont similaires : le 20 juillet reste le jour avec le maximum de naissances et trois jours fin septembre figurent en 4[ème], 5[ème] et 6[ème] position. Le nombre moyen de naissances désaisonnalisé (nd) par jour (j) est égal à :

$$nc = \sum_{j=1}^{7} \frac{n}{10} \times \frac{m_j}{s_j}$$

avec
n : nombre moyen de naissances par jour
m : nombre de jour j de la semaine sur la période
s : sur ou sous-natalité lors du jour j (données de la figure 2a)

De 2015 à 2024, il y a eu trois 29 février, qui ont eu lieu un lundi en 2016, un samedi en 2020 et un jeudi en 2024. La sous-natalité le 29 février est le rapport entre le nombre moyen des naissances le 29 février sur la période et le nombre moyen des naissances par jour lors des lundis de 2016, des samedis de 2020 et les jeudis de 2024.

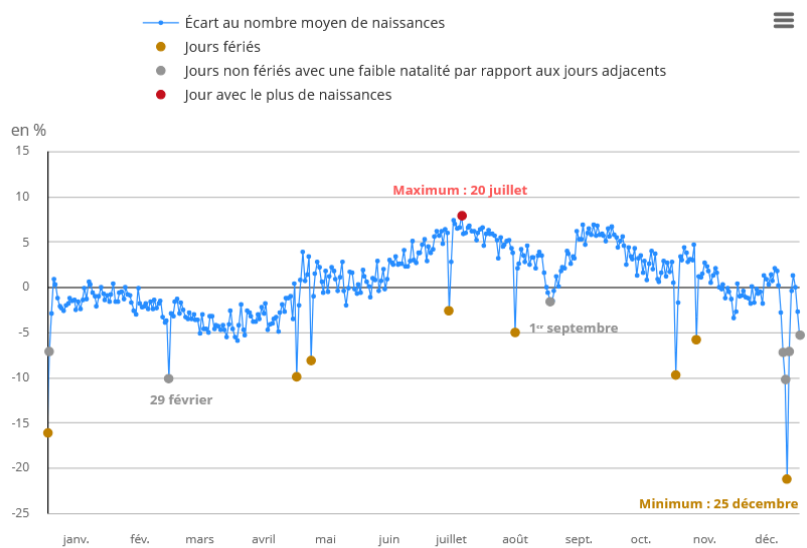

Figure 1b - Écart au nombre moyen de naissances par jour de 2015 à 2024, en corrigeant les effets des jours de la semaine

Notes : Seuls les jours fériés à date fixe peuvent être représentés. La sur ou sous-natalité par jour a été corrigée pour tenir compte du fait que chaque jour comporte un nombre différent de jours de la semaine sur la période. Par exemple, de 2015 à 2024, le 20 juillet a compté deux lundis, un mardi, deux mercredis, deux jeudis, un vendredi, deux samedis et aucun dimanche (**méthodes**).
Lecture : De 2015 à 2024, en corrigeant les effets des jours de la semaine, le nombre moyen de naissances le 25 décembre était inférieur de 21,2 % au nombre moyen de naissances par jour.
Champ : Naissances ayant eu lieu en France.
Source : Insee, statistiques d'état civil.

## Définitions

Une **vague de chaleur** désigne un épisode de températures nettement plus élevées que les normales pendant plusieurs jours (définition à l'échelle nationale) **[Météo France, 2025]**. Entre 1947 et 2024, Météo-France a comptabilisé 49 vagues de chaleur.

Une **canicule** désigne un épisode de températures élevées de jour comme de nuit sur une période prolongée (au moins 3 jours) qui est susceptible de constituer un risque sanitaire notamment pour les personnes fragiles ou surexposées (définition à l'échelle départementale) **[Météo France, 2025]**.

La catégorie des **professeurs des écoles, instituteurs et assimilés** inclut notamment :
- les enseignants de l'enseignement primaire,

- les enseignants non agrégés ou non certifiés de l'enseignement secondaire,
- les conseillers principaux d'éducation,
- les formateurs de formation continue,
- les éducateurs sportifs.

La catégorie ne comprend pas les enseignants agrégés ou certifiés de l'enseignement secondaire.

**Pour en savoir plus**

Météo France, Canicule, pic ou vague de chaleur ?, septembre 2025

Cinelli H., Lelong N., Le Ray C., Demiguel V., Lebreton É., Deroyon T., « Les naissances, le suivi à deux mois et les établissements. Situation et évolution depuis 2016 », Enquête nationale périnatale, octobre 2022.

Papon S., « En un siècle, le pic des naissances s'est décalé de l'hiver à l'été et s'est atténué », Insee Focus n°204, septembre 2020.

Touzet C., Les naissances baissent légèrement Bilan démographique 2018 à Mayotte, Insee Flash Mayotte n°91, septembre 2019.

Kadhel P., Costet N., Toto T., Janky E., Multigner L., "The annual carnival in Guadeloupe is associated with an increase in the number of conceptions and subsequent births nine months later: 2000 - 2011", mars 2017.

Rohrbasser J.-M., Régnier-Loilier A., « Y a-t-il une saison pour faire des enfants ? », Population et Sociétés n° 474, Ined, janvier 2011.

Régnier-Loilier A., Wiles-Portier E., "Choosing the Time of Year for Births: A Barely Perceptible Phenomenon in France." Population (English Edition, 2002-), vol. 65 n°1, 2010.

Leridon H., « L'espèce humaine a-t-elle un problème de fertilité ? », Population et Sociétés n° 471, octobre 2010.

Baubeau D., Buisson G., « La pratique des césariennes : évolution et variabilité entre 1998 et 2001 », Études et Résultats n°275, décembre 2003.